\documentclass[11pt,twoside]{article}
\usepackage{asp2010}
\usepackage{graphicx}

\resetcounters

\begin{document}

\title{Asymptotic Giant Branch Variables in NGC\,6822}
\author{Francois Nsengiyumva$^{1,2}$, Patricia A. Whitelock$^{1,2}$, 
Michael W. Feast$^{2,1}$ and John W. Menzies$^1$
\affil{$^1$ South African Astronomical Observatory, PO Box 9, 7935 Observatory,
South Africa}
\affil{$^2$ Astronomy, Cosmology and Gravitation Centre, Astronomy
Department, University of Cape Town, 7701 Rondebosch, South Africa}}

\begin{abstract}
Using multi-epoch $JHK_S$ photometry obtained with the 1.4-m
Japanese-South African Infrared Survey Facility at Sutherland we have
identified large numbers of AGB variables in NGC\,6822.
This paper uses 30 large amplitude variables,
with periods ranging from about 200 to 900 days, to provide a new
calibration for the period-luminosity relation.
\end{abstract}

\section{Observations and the Mira Period-Luminosity Relation}
This work is part of a large programme to examine AGB variable
stars in Local Group galaxies. 
NGC\,6822 is a Magellanic-type dwarf irregular with a central bar and an
extended stellar halo; it is the nearest isolated dwarf galaxy in the Local Group. 

 The observations were made with the SIRIUS camera on the IRSF at SAAO. 
Three fields were used to cover a total area of 
$7.5 \times 21.1$ arcmin$^2$ and 19 observations were made over a period 
of four years. Only a subset of the data has been analyzed in detail, from which we have
identified 30 Mira variables with $\Delta K>0.4$ mag and 9 SR variables with lower
amplitudes.

\begin{figure}
\includegraphics[]{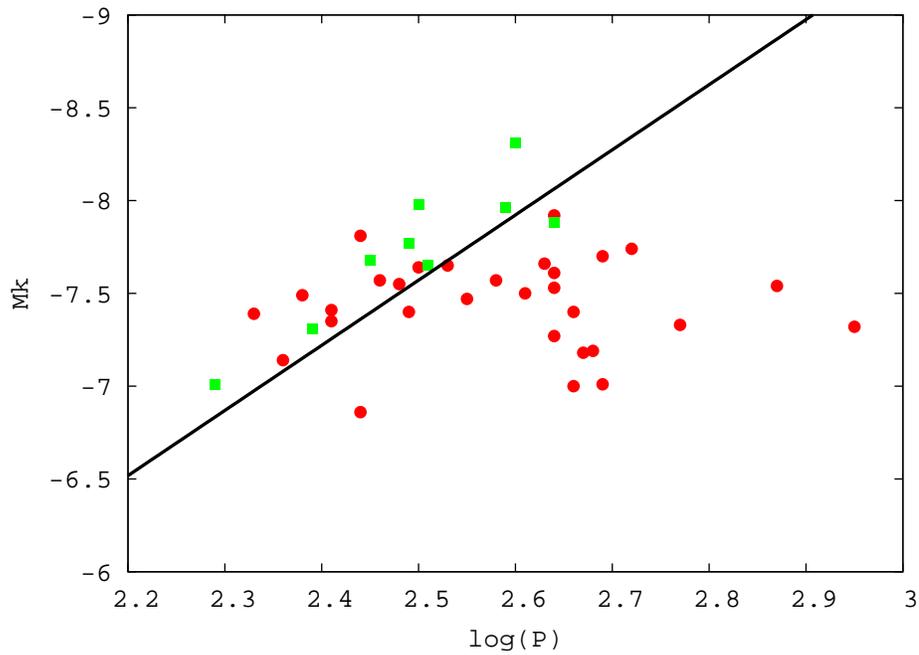}
\caption{PL($K$) relation for Miras (red circles) and SRs (green squares). The line
shows the LMC PL($K$) relation.}
\end{figure}

Fig.~1 shows the Miras and SRs on a period luminosity, PL($K$), relation, where
the absolute $K$ mags were determined on the assumption that the distance
modulus to NGC 6822 is 23.43 mag. It is clear that many of the 
Miras, particularly the redder and longer period ones, fall below the
predicted PL($K$) relation.  

\begin{figure}
\includegraphics[]{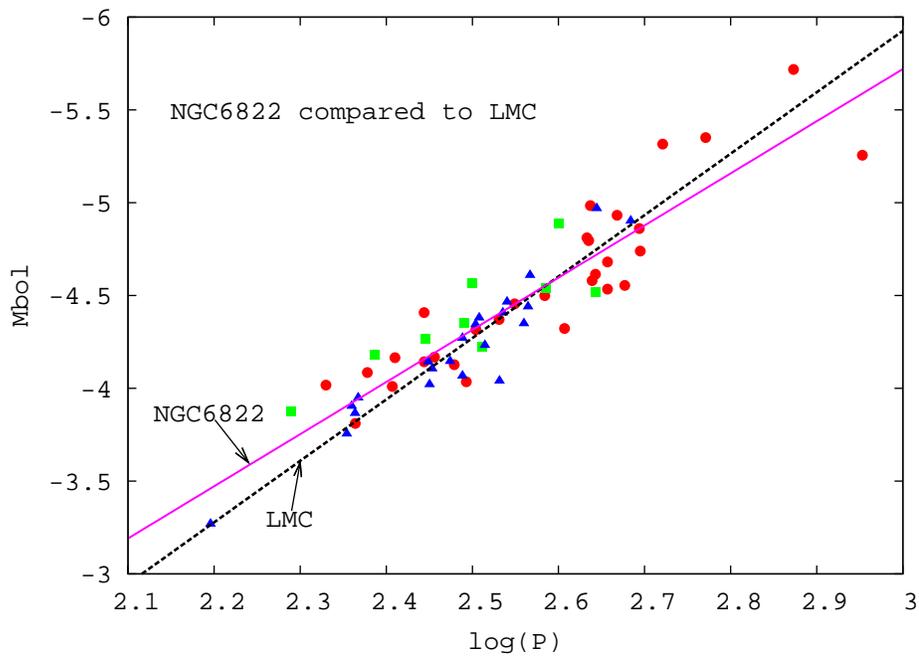}
\caption{Bolometric PL relation comparing the Miras in NGC\,6822 (red
circles) with the C-rich Miras in the LMC (blue triangles); SR variables in
NGC\,6822 are also shown as green squares. }
\end{figure}

Fig.~2 shows the bolometric PL relation and compares it with that obtained
for C-rich Miras in the LMC.  The agreement is good, so we can understand
the departure from the PL($K$) as a consequence of circumstellar extinction.  

\section{Conclusion}
Further work is required to fully characterize the other variables in NGC
6822 and to determine if they all are carbon-rich, as we assume here.  
This sample already provides a very good group with which to calibrate the
Mira PL relation beyond the Magellanic Clouds. We anticipate a good deal
of interest in this as the next generation of telescopes from the
ground and from space start to image ever more distant individual stars in
the infrared.


\end{document}